\documentclass[10 pt]{IEEEtran}

\usepackage[cmex10]{amsmath}
\usepackage{amssymb, amsfonts}
\usepackage{color,epsfig}

\usepackage{cite}
\newtheorem{definition}{Definition}
\newtheorem{theorem}{Theorem}

\newtheorem{remark}{Remark}
\newcommand{\remarkend}{ \IEEEQEDopen}
\newtheorem{lemma}{Lemma}
\newenvironment{myalign}{\allowdisplaybreaks \align}{\endalign}

\newcommand{\bA}{{\mathbf A}}
\newcommand{\cE}{\mathcal E}
\newcommand{\cA}{\mathcal A}
\newcommand{\cB}{\mathcal B}
\newcommand{\cC}{\mathcal C}
\newcommand{\cH}{\mathcal H}
\newcommand{\bH}{{\mathbf H}}
\newcommand{\cF}{\mathcal F}
\newcommand{\cG}{\mathcal G}
\newcommand{\cV}{\mathcal V}
\newcommand{\bZ}{{\mathbf Z}}
\newcommand{\bS}{{\mathbf S}}
\newcommand{\bI}{{\mathbf I}}
\newcommand{\bX}{{\mathbf X}}
\newcommand{\bO}{{\mathbf O}}
\newcommand{\bksl}{\backslash}

\begin{document}

\title{MIMO Multiple Access Channel with an Arbitrarily Varying
  Eavesdropper \thanks{This work was presented in part at the 49th
    Annual Allerton Conference on Communication, Control, and
    Computing, September, 2011. This work is supported in part by NSF
    Grant 0964362. A. Khisti's work was supported by an NSERC
    Discovery Grant. The ordering of authors is alphabetical.} }
\author{ \IEEEauthorblockN{Xiang He, Ashish
    Khisti\IEEEauthorrefmark{1}, Aylin
    Yener\IEEEauthorrefmark{2}} \\
  \IEEEauthorblockA{ \IEEEauthorrefmark{1}Dept. of Electrical and
    Computer Engineering,
    University of Toronto,  Toronto, ON, M5S 3G4, Canada\\
    \IEEEauthorrefmark{2} Electrical Engineering Department, The
    Pennsylvania
    State University, University Park, PA 16802\\
    \textit{xianghe@microsoft.com, akhisti@comm.utoronto.ca,
      yener@psu.edu}}}

\maketitle

\begin{abstract}
  A two-transmitter Gaussian multiple access wiretap channel with
  multiple antennas at each of the nodes is investigated. The channel
  matrices at the legitimate terminals are fixed and revealed to all
  the terminals, whereas the channel matrix of the eavesdropper is
  arbitrarily varying and only known to the eavesdropper. The secrecy
  degrees of freedom (s.d.o.f.) region under a strong secrecy
  constraint is characterized.  A transmission scheme that
  orthogonalizes the transmit signals of the two users at the intended
  receiver and uses a single-user wiretap code is shown to be
  sufficient to achieve the s.d.o.f.\ region.  The converse involves
  establishing an upper bound on a weighted-sum-rate expression. This
  is accomplished by using induction, where at each step one combines
  the secrecy and multiple-access constraints associated with an
  adversary eavesdropping a carefully selected group of sub-channels.
\end{abstract}

\section{Introduction}
\label{sec:introduction}
Information theoretic security was first introduced by Shannon in
\cite{shannonsecrecy}, which studied the problem of transmitting
confidential information in a communication system in the presence of
an eavesdropper with unbounded computational power. Since then, an
extensive body of work has been devoted to studying this problem for
different network models by deriving fundamental transmission rate
limits \cite{csiszar1978bcc, liang2009information, bloch2011physical}
and designing low-complexity schemes to approach these limits in
practice \cite{thangaraj2005alc, bellare2012polynomial}.

Secure communication using multiple antennas was extensively studied
as well, see e.g., \cite{li2007using, goel2008guaranteeing,
  khisti:isit, khisti:stm, khisti:stm2, liu2010multiple, ekrem-2009,
  kobayashi-2009,mukherjee-2011}. These works investigated efficient
signaling mechanisms using the spatial degrees of freedom provided by
multiple antennas to limit an eavesdropper's ability to decode
information.  The underlying information theoretic problem, the
multi-antenna wiretap channel, was studied and the associated secrecy
capacity was identified. We note that these works assumed that the
eavesdropper's channel state information is available either
completely or partially, although such an assumption may not be
justified in practice.

As a more pessimistic but stronger assumption, references
\cite{Ebrahimthesis, xiang2010arb, xiang2011globecom} study secrecy
capacity when the eavesdropper channel is arbitrarily varying and its
channel states are known to the eavesdropper only. Reference
\cite{xiang2010arb} studies the single-user Gaussian
multi-input-multi-output (MIMO) wiretap channel and characterizes the
secrecy degrees of freedom (s.d.o.f.).  The same paper extended the
single user analysis to the two user Gaussian MIMO multiple access
(MIMO-MAC) channel. This was possible only when all the legitimate
terminals have equal number of antennas, leaving the MIMO-MAC with
arbitrary number of antennas at the terminals an open problem.

Our main contribution is to fully characterize the s.d.o.f. region of
the two-transmitter MIMO MAC channel when the eavesdropper channel is
arbitrarily varying. We show that the s.d.o.f. region can be achieved
by a scheme that orthogonalizes the transmit signals of the two users
at the intended receiver. Moreover, it suffices to use a single-user
wiretap channel code~\cite{xiang2010arb} and no coordination between
the users is necessary except for synchronization and sharing the
transmit dimensions. To establish the optimality of this scheme, our
converse proof decomposes the MIMO MAC channel into a set of parallel
and independent channels using the generalized singular value
decomposition (GSVD).  A set of eavesdroppers, each monitoring a
subset of links, is selected using an induction procedure and the
resulting secrecy constraints are combined to obtain an upper bound on
a weighted sum-rate expression. The outer bound matches the achievable
rate in terms of the s.d.o.f.\ region, thus settling the open problem
raised in \cite{xiang2010arb} for the case of two transmitters.

Interestingly, the s.d.o.f.\ region remains open for this model when
the eavesdropper channel is perfectly known to all terminals. A
significant body of literature already exists on this problem, see
e.g.,~\cite{ender-special, liang2008macwt, ersen2008allerton,
  dimakis2010allerton}. If the channel model has real inputs and
outputs, Gaussian signaling is in general suboptimal and user
cooperating strategies as well as signal alignment techniques are
necessary~\cite{xiangITstructure}. In~\cite{goel2010modelingisit} it
is established that s.d.o.f.\ of $1/2$ is achievable using real
interference alignment for almost all configurations of channel
gains. If the channel model has complex inputs and outputs, it is
shown in \cite[Section 5.16]{xiangthesis} that in general s.d.o.f.\ of
$1/2$ is achievable using asymmetric Gaussian signaling. In contrast,
the best known upper bound on the s.d.o.f. of individual rates is
$2/3$ for both cases, established in \cite[Section 5.5]{xiangthesis}.

The remainder of this paper is organized as follows. In
Section~\ref{sec:model}, we describe the system model. The main result
is stated as Theorem~\ref{thm:mainresult} in
Section~\ref{sec:mainresult}. The proof of the theorem is divided into
two parts.  First, we establish the result for the case of parallel
channels in Section~\ref{sec:Parallel}. Subsequently, in
Section~\ref{sec:GeneralMIMO} we establish the result for the general
case by decomposing the MIMO-MAC channel into a set of independent
parallel channels.  Such a reduction is used both in the proof of the
converse as well as the coding scheme. Section~\ref{sec:conclusion}
concludes the paper.

We use the following notation throughout the paper: For a set $\cA$,
$V_{i,\cA}$ and $V_{\cA}$ denote the set of random variables
$\{V_{i,j}, j \in \cA\}$ and $\{V_j, j \in \cA\}$
respectively. $\{\delta_n\}$ denotes a non-negative sequence of $n$
that converges to $0$ when $n$ goes to $\infty$. We use bold
upper-case font for matrices and vectors and lower-case font for
scalars. The distinction between matrices and vectors will be clear
from the context. For a set $\cA$, $|\cA|$ denotes its cardinality and
a short hand notation $x^n$ is used for the sequence $\{x_1,
x_2,\ldots, x_n\}$. $\phi$ denotes the empty set.

\section{System Model}
\label{sec:model}
\begin{figure}[t]
  \centering 
\includegraphics[scale=0.65]{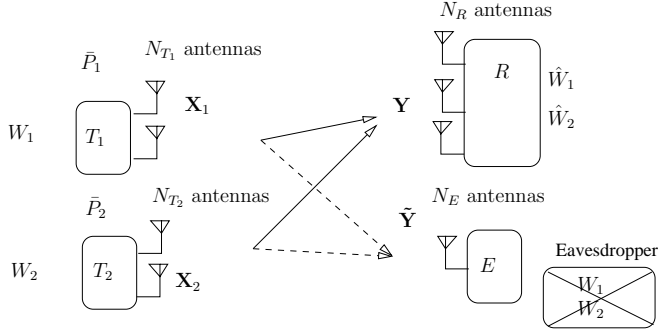}
  \caption{The MIMO MAC wiretap channel where $N_{T_1}=N_{T_2}=2$,
    $N_R=3$, $N_E=1$.}
  \label{fig:2usermac}
\end{figure}
As shown in Figure~\ref{fig:2usermac}, we consider a discrete-time
channel model where two transmitters communicate with one receiver in
the presence of an eavesdropper. We assume transmitter $i$ has
$N_{T_i}$ antennas, $i=1,2$, the legitimate receiver has $N_R$
antennas whereas the eavesdropper has $N_E$ antennas.  The channel
model is given by
\begin{myalign}
  & \mathbf{Y}(i) = \sum\limits_{k = 1}^2 {\mathbf{H}_k \mathbf{X}_k }(i)  + \mathbf{Z}(i)  \label{eq:mimomac}\\
  & \mathbf{\tilde Y} (i) = \sum\limits_{k = 1}^2 {\mathbf{\tilde H}_k
    (i) \mathbf{X}_k (i) }\label{eq:eavesdropperchannel}
\end{myalign}
where $i\in \{1,\ldots, n\}$ denotes the time-index, $\mathbf{H}_k,
k=1,2,$ are channel matrices and $\mathbf{Z}$ is the additive Gaussian
noise observed by the intended receiver, which is composed of
independent rotationally invariant complex Gaussian random variables
with zero mean and unit variance.  The sequence of eavesdropper
channel matrices $\{\mathbf{\tilde H}_k(i), k=1,2\},$ is an arbitrary
sequence of length $n$ and only revealed to the eavesdropper.  In
contrast, $\mathbf{H}_k, k=1,2$ are revealed to both the legitimate
parties \textit{and} the eavesdropper(s). We assume $N_E$, the number
of eavesdropper antennas, is known to the legitimate parties and the
eavesdropper.

User $k$, $k=1,2$, wishes to transmit a confidential message $W_k$,
$k=1,2$, to the receiver over $n$ channel uses, while both messages,
$W_1$ and $W_2$, must be kept confidential from the eavesdropper. We
use $\gamma$ to index a specific sequence of $\{\mathbf{\tilde
  H}_k(i), k=1,2\}$ over $n$ channel uses and use $\mathbf{\tilde
  Y}_{\gamma}^n$ to represent the corresponding channel outputs for
$\mathbf{\tilde Y}^n$. The \textit{strong} secrecy constraint is
\cite{xiang2010arb}:
\begin{myalign}
  \mathop {\lim }\limits_{n \to \infty } I\left( {W_1 ,W_2
      ;\mathbf{\tilde Y}_{\gamma}^n } \right) = 0, \quad \forall
  \gamma \label{eq:secrecy2}
\end{myalign}
where the convergence  must be uniform over $\gamma$.  The average power constraints for the two users are given by
\begin{myalign}
  \lim_{n \to \infty}\frac{1}{n}\sum_{i=1}^n |\mathbf{X}_k(i)|^2 \le \bar P_k,\quad k=1,2.\label{eq:power2}
\end{myalign}
The secrecy rate for user $k$, $R_{s,k}$, is defined as
\begin{myalign}
  R_{s,k}=\lim_{n \to \infty} \frac{1}{n} H(W_k),\quad k=1,2. \label{eq:def_R_e_k}
\end{myalign}
such that $W_k$ can be reliably decoded by the receiver, and
\eqref{eq:secrecy2} and \eqref{eq:power2} are satisfied.

We define the secrecy degrees of freedom as:
\begin{myalign}
  \left\{(d_1,d_2): d_k=\limsup_{\bar P_1=\bar P_2=\bar P \to \infty}\frac{ R_{s,k}}{\log_2 \bar
    P}, \quad k=1,2\right\}\label{eq:def_sdof_region}
\end{myalign}

\section{Motivation}
\label{subsec:motivtation}
\begin{figure}[t]
  \centering 
\includegraphics[scale=0.7]{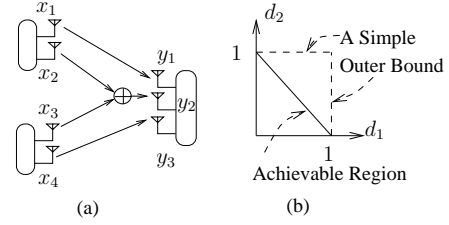}
  \caption{(a) A special case of MIMO MAC wiretap channel where $N_{T_1}=N_{T_2}=2$,
    $N_R=3$, $N_E=1$, (b) Comparison between achievable
    s.d.o.f. region and a simple outer bound derived by considering
    one eavesdropper at a time.}
  \label{fig:motivatingexample}
\end{figure}
Before stating the main result, we illustrate the main difficulty in
characterizing the s.d.o.f. region through a simple example.  As
illustrated in Figure~\ref{fig:motivatingexample}(a), in this example,
each transmitter has 2 antennas and the intended receiver has 3
antennas, while the eavesdropper has only 1 antenna. Let $x_1, x_2,
x_3, x_4$ denote the transmitted signals from the two users and $y_1,
y_2, y_3$ denote the signals observed by the intended the
receiver. And the main channel is given by
\begin{myalign}
&  y_1=x_1+z_1, \quad  y_3=x_4+z_3\\
&  y_2=x_2+x_3+z_2
\end{myalign}
where $z_i, i=1,2,3$ denote additive channel noise. As shown in
\cite{xiang2010arb}, a secrecy degree of freedom $\min(N_{T_k},
N_R)-N_E=1$ is achievable for a user if the other user remains
silent. Time sharing between these two users lead to the following
achievable s.d.o.f. region:
\begin{myalign}
  d_1+d_2 \le 1, \quad d_k \ge 0, \quad k=1,2 \label{eq:motivatingexampleachievableregion}
\end{myalign}

For the converse, we begin by considering a simple upper bound, which
reduces each channel to a single-user MIMO wiretap channel.  First, by
revealing the signals transmitted by user $2$ to the intended receiver
and assuming that the eavesdropper monitors either $x_1$ or $x_2$ we
have that $d_1 \le 1$.  Similarly we argue that $d_2 \le 1$. To obtain
an upper bound on the sum-rate we let the two transmitters to
cooperate and reduce the system to a $3 \times 3$ MIMO link. The
s.d.o.f. of this channel~\cite{xiang2010arb} yields $d_1 + d_2 \le 2$.
This outer bound, illustrated in
Figure~\ref{fig:motivatingexample}(b), does not match with the
achievable region given by
\eqref{eq:motivatingexampleachievableregion}.

As we shall show in Theorem~\ref{thm:mainresult},
\eqref{eq:motivatingexampleachievableregion} is indeed the
s.d.o.f. capacity region and hence a new converse is necessary to
prove this result. Our key observation is that the above upper bound
only considers one eavesdropper at a time in deriving each of the
three bounds.  For example, when deriving $d_1 \le 1$, we assume there
is only one eavesdropper which is monitoring either $x_1$ or
$x_2$. When deriving $d_2 \le 1$, we assume there is only one
eavesdropper which is monitoring either $x_3$ or $x_4$.  Similarly
when deriving $d_1 + d_2 \le 2$ we again assume that there is one
eavesdropper on either of the links.  As we shall discuss below, a
tighter upper bound is possible to find if we consider the
simultaneous effect of two eavesdroppers.

In our system model, there are infinitely many possible eavesdroppers,
each corresponding to a different channel state sequence. The
challenge is to find out a finite number of eavesdroppers, whose joint
effect leads to a tight converse. Our choice of eavesdroppers is based
on the following intuition: When an eavesdropper chooses which links
to monitor, it should give precedence to those links over which only
one user can transmit. This is because these links are the major
contributor to the sum s.d.o.f. $d_1+d_2$ since they are dedicated
links to a certain user. Based on this intuition, we consider the
following two eavesdroppers: one monitors $y_1$ for $W_1$ and the
other monitors $y_3$ for $W_2$. As we shall show later in
Lemma~\ref{lemma:case1_R1R2_bound}, the first eavesdropper implies the
following upper bound on $R_1$:
\begin{myalign}
n(R_1  - \delta _n ) \le I\left( {x_2^n ;y_2^n |y_1^n ,x_{\left\{
        {3,4} \right\}}^n } \right) \label{eq:R1_motivatingexample}
\end{myalign}
and the second eavesdropper implies the following upper bound on
$R_2$:
\begin{myalign}
n(R_2  - \delta _n ) \le I\left( {y_1^n ,x_{\left\{ {3,4} \right\}}^n
    ;y_2^n } \right) \label{eq:R2_motivatingexample}
\end{myalign}
Their joint effect can be captured by adding
\eqref{eq:R1_motivatingexample} and \eqref{eq:R2_motivatingexample}
\cite{ashishcompound}, which lead to:
\begin{myalign}
  n(R_1 + R_2 - 2\delta _n ) \le I\left( {x_2^n ,y_1^n ,x_{\left\{
          {3,4} \right\}}^n ;y_2^n } \right)
\end{myalign}
Since there is only one term, which is $y_2^n$, at the right side of
the mutual information $I\left( {x_2^n ,y_1^n ,x_{\left\{ {3,4}
      \right\}}^n ;y_2^n } \right)$, we observe the sum s.d.o.f. can
not exceed $1$, thereby justifying that
\eqref{eq:motivatingexampleachievableregion} is indeed the largest
possible s.d.o.f. region for Figure~\ref{fig:motivatingexample}(a).

As captured by \eqref{eq:R1_motivatingexample} and
\eqref{eq:R2_motivatingexample}, a simultaneous selection of two
different eavesdroppers for the two users reduces the effective signal
dimension at the receiver from three to one, thus leading to a tighter
converse. As we shall show later in
Section~\ref{subsec:converse-parallel-case-2}, in generalizing this
example we are required to systematically select a sequence of
eavesdroppers using an induction procedure.

\section{Main Result}
\label{sec:mainresult}
In this section, we state the main result of this work. To express our
result, we define $r_t$ as the rank of $\mathbf{H}_t, t=1,2$ and $r_0$
as the rank of $[~\mathbf{H}_1~|~ \mathbf{H}_2~]$. We will refer to
$r_t$ as the number of transmit dimensions at user $t=1,2$ and $r_0$
as the number of dimensions at the receiver.

\begin{theorem}
  The secrecy degrees of freedom region of the MIMO multiple access
  channel with arbitrarily varying eavesdropper channel is given by
  the convex hull of the following five points of $(d_1,d_2)$:
  \begin{myalign}
&    p_0=(0, 0)\\
&    p_1=\left(\left[r_1-N_E\right]^+, 0\right)\\
&    p_2=\left(0, \left[r_2-N_E\right]^+\right)\\
&    p_3=\left(\left[r_1-N_E\right]^+, \left[r_0-r_1-N_E\right]^+\right) \label{eq:p3}\\
&    p_4=\left(\left[r_0-r_2-N_E\right]^+, \left[r_2-N_E\right]^+\right) \label{eq:p4}
  \end{myalign}
where we use $[x]^+ \stackrel{\Delta}{=}\max\{x,0\}$.
\label{thm:mainresult}
\end{theorem}

\begin{figure}[t]
  \centering 
\includegraphics[scale=0.7]{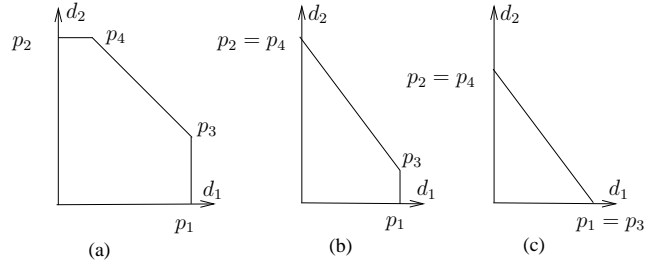}
 \caption{The secrecy degrees of freedom (s.d.o.f.) region in
  Theorem~\ref{thm:mainresult}: (a) $0 \le N_E \le
  \min\{r_0-r_1,r_0-r_2\}$, (b) $\min\{r_0-r_1,r_0-r_2\} \le N_E \le
  \max\{r_0-r_1,r_0-r_2\}$, (c) $\max\{r_0-r_1,r_0-r_2\} \le N_E$}
  \label{fig:dofregion}
\end{figure}

Fig.~\ref{fig:dofregion} illustrates the structure of the
s.d.o.f. region as a function of the number of eavesdropping
antennas. In Fig.~\ref{fig:dofregion}~(a) we have $N_E \le
\min(r_0-r_1, r_0-r_2)$. In this case the s.d.o.f. region is a
polymatroid (see e.g.,~\cite[Definition 3.1]{tse1998multiaccess})
described by $d_i \le r_i - N_E$ and $d_1 + d_2 \le r_0 - 2N_E$.  
Fig.~\ref{fig:dofregion}~(b) illustrates the shape of the s.d.o.f.\
region when $ \min\{r_0-r_1, r_0-r_2\} \le N_E \le \max\{r_0-r_1,
r_0-r_2\}$. In Fig.~\ref{fig:dofregion}~(b), without loss of
generality, we assume $r_1 < r_2$ and the s.d.o.f.\ region is bounded
by the lines $d_i \ge 0$, $d_1 \le r_1-N_E$ and
\begin{myalign}
&(r_1 + r_2 - r_0){d_1} + (r_1-N_E){d_2} \nonumber\\
&\le (r_1-N_E) \times (r_2-N_E).\label{eq:dof_region_b}
\end{myalign}
When $\min(r_1,r_2) > N_E \ge \max(r_0-r_1, r_0-r_2)$, the s.d.o.f.\
region, as illustrated in Fig.~\ref{fig:dofregion}~(c) is bounded by
$d_i \ge 0$ and the line 
\begin{equation}
\frac{d_1}{r_1-N_E} +\frac{d_2}{r_2-N_E} \le 1.
\label{eq:dof_region_c}
\end{equation}

\begin{figure}[t]
  \centering 
\includegraphics[scale=0.85]{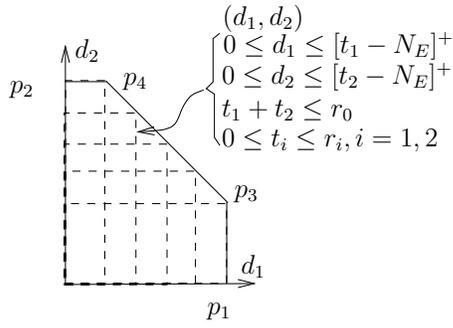}
  \caption{Interpretation of the s.d.o.f. region as a convex hull of
    rectangles: $(d_1 ,d_2 ):0 \le d_i \le [t_i - N_E ]^ + ,i = 1,2$,
    where $t_i$ is the number of degrees of freedom occupied by user
    $i$. To achieve reliable transmission, we must have
    \eqref{eq:t_i_constraint} and \eqref{eq:sum_t_i_constraint}.}
  \label{fig:interpretation}
\end{figure}

The s.d.o.f. region in Theorem~\ref{thm:mainresult} allows the
following simple interpretation: The region can be expressed as a
convex hull of a set of rectangles shown by
Figure~\ref{fig:interpretation} (illustrated for
Figure~\ref{fig:dofregion}~(a)). Each rectangle is parameterized by
the dimensions of the subspace occupied by the transmission signals
from the two users, denoted by $(t_1,t_2)$, where $t_i$ indicates the
dimension of user $i$, $i=1,2$. Then in order for the signals from
both transmitters to be received reliably by the receiver, we must
have
\begin{myalign}
&  t_1 + t_2 \le r_0 \label{eq:t_i_constraint}\\ 
&  0 \le t_i \le r_i ,i = 1,2 \label{eq:sum_t_i_constraint}
\end{myalign}
Each user then transmits confidential messages with $0\le d_i\le
[t_i-N_E]^+$ over the available $t_i$ dimensions, where the $-N_E$
term is an effect of the secrecy constraint \eqref{eq:secrecy2}.

It is clear that $p_3, p_4$ given by \eqref{eq:p3} and \eqref{eq:p4}
are in one of these rectangles. Hence the convex hull of these
rectangles yields the s.d.o.f. region stated in
Theorem~\ref{thm:mainresult}.

\section{Proof for the Parallel Channel Model }
\label{sec:Parallel}
 \begin{figure}[t]
  \centering 
\includegraphics[scale=0.7]{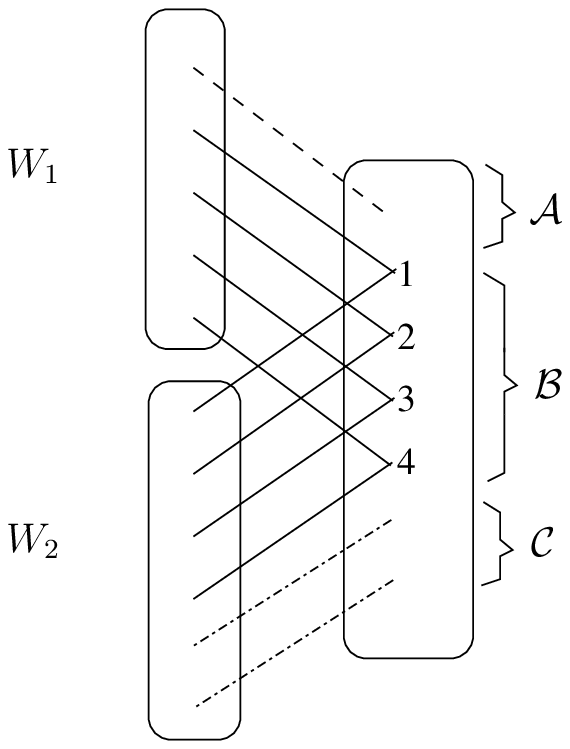}
 \caption{Definition of the set $\cA, \cB, \cC$, where $|\cB|=4$.}
  \label{fig:setA,B,C}
\end{figure}
In this section, we establish Theorem~\ref{thm:mainresult} for the
case of parallel channels.  As illustrated in Fig.~\ref{fig:setA,B,C},
the receiver observes
\begin{align}
y_i = x_{1i} + z_i, \quad i \in \cA,\label{eq:output_A}\\
y_i = x_{1i} + x_{2i} + z_i, \quad i \in \cB,\label{eq:output_B}\\
y_i = x_{2i} + z_i, \quad i \in \cC,\label{eq:output_C}
\end{align}where the noise random variables across the sub-channels are independent and each is
distributed according to ${\mathcal{CN}}(0,1)$ and $\{x_{1i}\}_{i \in \cA \cup \cB}$ and $\{x_{2i}\}_{i \in \cB \cup \cC}$
denote the transmit symbols of user 1 and user 2 respectively.

The parallel channel model is a special case of~\eqref{eq:mimomac} with
\begin{align}
\bH_1\!=\!\left[ \begin{array}{*{20}c}
   {\bI}_{|\cA|}  & {} & {}  \\
   {} & {\bI_{ |\cB|} } & {} \\
      {} & { } & {\bO_{|\cC|}} 
\end{array} \right]\!,~
\bH_2 \!= \! \left[ \begin{array}{*{20}c}
   {\bO}_{|\cA|}  & {} & {}  \\
   {} & {\bI_{ |\cB|} } & {} \\
      {} & { } & {\bI_{|\cC|}} 
\end{array} \right]\!,  \label{eq:H1H2}
\end{align}
where $\bI_{|\cA|}$, $\bI_{|\cB|}$ and $\bI_{|\cC|}$ denote the
identity matrices of size $|\cA|$, $|\cB|$ and $|\cC|$ respectively,
and $\bO_{|\cA|}$ and $\bO_{|\cB|}$ denote the matrices, all of whose
entries are zeros.  Note that we do not make any assumption on the
eavesdropper's channel model~\eqref{eq:eavesdropperchannel}.

\subsection{Achievability}
\label{subsec:achievable-parallel}
It suffices to establish the achievability of points $p_3$ and $p_4$
in~\eqref{eq:p3} and~\eqref{eq:p4} respectively. The rest of the
region follows through time-sharing between these points. Note that
for the proposed parallel channel model
\begin{align}
p_3 &= \left( \left[|\cA| + |\cB|- N_E\right]^+, \left[|\cC|- N_E \right]^+\right)\label{eq:p3_parallel}\\
p_4 &= \left( \left[|\cA| - N_E\right]^+, \left[|\cB| + |\cC|- N_E \right]^+\right)\label{eq:p4_parallel}
\end{align}

To prove the achievability of $p_3$ we restrict user $2$ to transmit
only on the last $|\cC|$ components of in \eqref{eq:output_C} and
allow user $1$ to transmit over all of the components of $\cA \cup
\cB$ in \eqref{eq:output_A} and~\eqref{eq:output_B}. Note that in this
case, the signals of these two users do not interfere with each other
at the intended receiver. From \cite{xiang2010arb}, user $1$ can
transmit $W_1$ such that $d_1=[|\cA| + |\cB|-N_E]^+$ and
\begin{myalign}
  \lim_{n \to \infty} I(W_1; \mathbf{\tilde H}_1^n \mathbf{X}_1^n)=0\label{eq:W1issecure}
\end{myalign}
and user $2$ can transmit $W_2$ such that $d_2=[|\cC|-N_E]^+$ and
\begin{myalign}
  \lim_{n \to \infty} I(W_2; \mathbf{\tilde H}_2^n \mathbf{X}_2^n)=0\label{eq:W2issecure}
\end{myalign}
where we use  $\mathbf{\tilde H}_k^n \mathbf{X}_k^n$ to denote
the sequence $\{\mathbf{\tilde H}_k(i) \mathbf{X}_k(i), i=1,...,n\}$.  
Furthermore since $(W_1,\bX_1^n$ is independent of $(W_2, \bX_2^n)$ we have that
\begin{myalign}
  \lim_{n \to \infty} I(W_1; \mathbf{\tilde H}_1^n \mathbf{X}_1^n,\mathbf{\tilde H}_2^n \mathbf{X}_2^n)=0\label{eq:W1issecure_2}\\
  \lim_{n \to \infty} I(W_2; \mathbf{\tilde H}_1^n \mathbf{X}_1^n,\mathbf{\tilde H}_2^n \mathbf{X}_2^n)=0\label{eq:W2issecure_2}
\end{myalign}
which imply:
\begin{myalign}
& I\left( {W_1 ;\mathbf{\tilde H}_1^n \mathbf{X}_1^n, \mathbf{\tilde H}_2^n \mathbf{X}_2^n|W_2 } \right) \notag\\ 
  \le& I\left( {W_1 ;W_2 ,\mathbf{\tilde H}_1^n \mathbf{X}_1^n,\mathbf{\tilde H}_2^n \mathbf{X}_2^n} \right) \\ 
  =& I\left( {W_1 ;\mathbf{\tilde H}_1^n \mathbf{X}_1^n,\mathbf{\tilde H}_2^n \mathbf{X}_2^n} \right) + I\left( {W_1 ;W_2 |\mathbf{\tilde H}_1^n \mathbf{X}_1^n, \mathbf{\tilde H}_2^n \mathbf{X}_2^n} \right) \\ 
  \le& I\left( {W_1 ;\mathbf{\tilde H}_1^n \mathbf{X}_1^n, \mathbf{\tilde H}_2^n \mathbf{X}_2^n} \right) + I\left( {W_1 ,\mathbf{\tilde H}_1^n \mathbf{X}_1^n;W_2, \mathbf{\tilde H}_2^n \mathbf{X}_2^n } \right) \\ 
  =& I\left( {W_1 ;\mathbf{\tilde H}_1^n \mathbf{X}_1^n} \right) 
\end{myalign}
where the last step follows from the fact that $(W_2,\mathbf{\tilde H}_2^n \mathbf{X}_2^n)$ is independent
from $(W_1, \mathbf{\tilde H}_1^n \mathbf{X}_1^n)$. Therefore
\eqref{eq:W1issecure} implies
\begin{myalign}
  \lim_{n \to \infty} I\left( {W_1 ;\mathbf{\tilde H}_1^n \mathbf{X}_1^n, \mathbf{\tilde H}_2^n \mathbf{X}_2^n|W_2 } \right)=0.\label{eq:W1issecuregivenW2}
\end{myalign}
Adding \eqref{eq:W1issecuregivenW2} and \eqref{eq:W2issecure_2}, we
obtain
\begin{myalign}
  \lim_{n \to \infty} I\left( {W_1, W_2 ;\mathbf{\tilde H}_1^n \mathbf{X}_1^n, \mathbf{\tilde H}_2^n \mathbf{X}_2^n } \right)=0\label{eq:W1issecuregivenW2b}
\end{myalign}
and the secrecy constraint~\eqref{eq:secrecy2} follows from the
data-processing inequality. Also, since the convergence in $n$ in
\eqref{eq:W1issecure_2} and \eqref{eq:W2issecure_2} is uniform
\cite{xiang2010arb}, the convergence in \eqref{eq:W1issecuregivenW2b}
and hence in \eqref{eq:secrecy2} is uniform as well. Hence we have
proved the point $p_3$ is achievable.

The achievability of $p_4$ is proved by repeating the
argument above by exchanging user $1$ with user $2$.

\begin{remark}
\label{remark:independentsecrecyguarantee}
As is evident from \eqref{eq:W1issecuregivenW2b}, the secrecy
guarantee achieved by one user is not affected by the transmission
strategy of the other user. \remarkend
\end{remark}

\subsection{Converse : $N_E \le \min(|\cA|,|\cC|)$}
\label{subsec:converse-parallel-case-1}
We need to show that the s.d.o.f. region is contained within 
\begin{align}
d_1 &\le |\cA| + |\cB|  - N_E \label{eq:case1bnd1} \\
d_2 &\le |\cC| + |\cB|  - N_E \label{eq:case1bnd2}\\
d_1 + d_2 &\le |\cA| + |\cB|  + |\cC| - 2N_E \label{eq:case1bnd12}
\end{align}
Since~\eqref{eq:case1bnd1} and~\eqref{eq:case1bnd2} directly follow
from the single user case in \cite{xiang2010arb}, we only need to
show~\eqref{eq:case1bnd12}.

Let $\cE_k$ be the set of links such that an eavesdropper  is monitoring
for $W_k$, $k=1,2$. $|\cE_1|=|\cE_2|=N_E$. $\cA \supseteq \cE_1$, $\cC \supseteq
\cE_2$. We establish the following upper bound on the achievable rate pairs.
\begin{lemma}
  \begin{myalign}
    &   n (R_{s,1}-\delta_n) \le  I\left( { X_{1, \cA\backslash \cE_1}^n; Y_{\cA\backslash  \cE_1 }^n } \right) + I\left( {X_{1,\cB}^n  ;Y_\cB^n } | M\right) \label{eq:case1_R1_bound}\\
    & n (R_{s,2}-\delta_n) \le I(X_{2, \cC \backslash \cE_2}^n;
    Y_{\cC\backslash \cE_2}^n) + I\left( M ;Y_\cB^n
    \right) \label{eq:case1_R2_bound}
  \end{myalign}
where $M=  \left(Y_{1, \cA}^n ,X_{2,\cB \cup \cC}^n\right)$.
\label{lemma:case1_R1R2_bound}
\end{lemma}
\begin{IEEEproof}
The proof is provided in Appendix~\ref{app:R1R2_case1}.
\end{IEEEproof}
The proof is completed upon adding
\eqref{eq:case1_R1_bound} and \eqref{eq:case1_R2_bound} so that
\begin{myalign}
&  n(R_{s,1}+R_{s,2}-2\delta_n) \nonumber\\
&\le (X_{1,\cA\bksl\cE_1}^n; Y_{1,\cA\bksl\cE_1}^n)+ I(X_{2,\cC\bksl\cE_2}^n;Y_{2,\cC\bksl\cE_2}^n)\nonumber\\
  &\quad +I(M, X_{1,\cB}^n;Y_{\cB}^n)
\end{myalign}
and using
\begin{myalign}
&  d\left( \frac{1}{n} I( { X_{\cA\backslash \cE_1}^n; Y_{\cA\backslash  \cE_1 }^n }
  )\right) \le |\cA|-N_E\\
&  d\left( \frac{1}{n}I(X_{\cC \backslash \cE_2}^n; Y_{\cC\backslash \cE_2}^n) \right) \le |\cC|-N_E\\
&  d\left( \frac{1}{n}I(M, X_{1, \cB}^n; Y_\cB^n) \right) \le |\cB|
\end{myalign}
where $d(x) \stackrel{\Delta}{=}\lim_{P\rightarrow\infty}\frac{x(P)}{\log_2 P}$ characterizes the pre-log scaling of $x$ with respect to $P$. 

\subsection{Converse : $N_E >\max(|\cA|,|\cC|)$}
\label{subsec:converse-parallel-case-2}
\begin{figure}[t]
  \centering 
\includegraphics[scale=0.7]{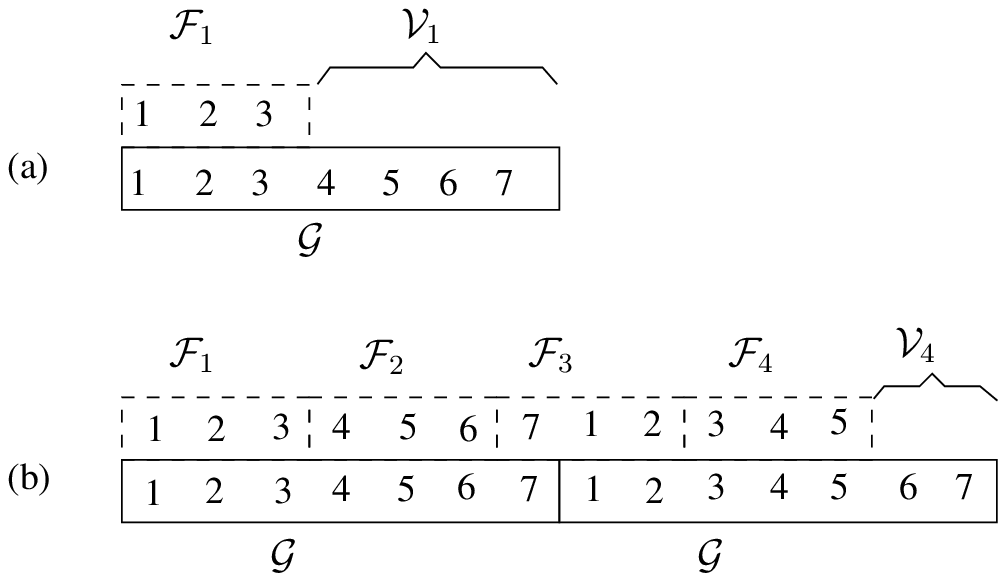}
  \caption{The set $\cF_k$, $\cG$, and $\cV_{k}$ when $|\cF|=3$, $|\cG|=7$
    and $|\cB|=8$. (a) Case I, $i=1$, $c_1=1$. (b) Case II, $i=4$, $\cH_5=\{1\}$,
    $\cF_5=\{6,7,1\}$, $\cV_5=\{2, 3, 4, 5, 6, 7\}$, $c_4=2$, $c_5=3$. }
  \label{fig:macFG}
\end{figure}

Without loss of generality, we assume $|\cC| \ge |\cA|$. Let $\cE_k$ be the
set of links such that an eavesdropper is monitoring for $W_k$,
$k=1,2$. Let $|\cE_1|=|\cE_2|=N_E$, $\cA \subset \cE_1$, and $\cC \subset \cE_2$.

Define the set $\cF, \cG$ such that $\cF = \cB \backslash \cE_1$, $\cG=\cB \backslash \cE_2$. Since $|\cC| \ge
|\cA|$, we have $|\cG| \ge |\cF|$.

Then Theorem~\ref{thm:mainresult} reduces to $d_k \ge 0, k=1,2$ and
\begin{myalign}
&  |\cG| d_1+ |\cF| d_2 \le |\cF| \times |\cG| \label{eq:converse_case2_parallel}
\end{myalign}which we now show. We first introduce the following lemma:
\begin{lemma}
For any choice of $   \cF \subseteq \cB$ and $\cG \subseteq \cB$ with
appropriate cardinalities the rates $R_{s,1}$ and $R_{s,2}$ are upper bounded
by
  \begin{myalign}
&   n (R_{s,1}-\delta_n) \le  I\left( {X_{1,\cF}^n;  Y_{\cF}^n  } | M, X_{1, \cB \backslash \cF}^n\right) \label{eq:R1bnd_case2_alt}\\
&   n (R_{s,2}-\delta_n) \le I\left( M ,X_{1, \cB \backslash \cG}^n ;Y_\cG^n \right) \label{eq:R2bnd_case2_alt}
  \end{myalign}
 where $M = \left\{Y_{1,\cA}^n ,X_{2,\cB \cup \cC}^n \right\}$.
\label{lemma:R1R2_case2}
\end{lemma}
\begin{IEEEproof}
The proof is provided in Appendix~\ref{app:R1R2_case2}.
\end{IEEEproof}

For the remainder of the proof we assume without loss of generality
that $\cB = \{1,\ldots, |\cB|\}$.  We fix $\cG = \{1,\ldots, |\cG| \}$
while choosing $|\cG|$ different sets of $|\cF|$ elements: $\cF_1,
\ldots, \cF_{|\cG|}$, the sets $\cV_0, \ldots, \cV_{|\cG|}$ and a
sequence of $c_i$ in the following recursive manner.
\begin{definition}
  Let $\cV_0= \cG$, $c_0=1$.  For $i \ge 1$ recursively construct
  $\cF_i$ as follows.
\begin{enumerate}
\item {\bf Case I: $|\cV_{i-1}| \ge |\cF|$} 

  Let $\cF_{i} = \{\cV_{i-1}(1), \ldots, \cV_{i-1}(|F|)\}$, where
  $\cV_{i-1}(k)$ denotes the $k$th smallest element in $\cV_{i-1}$.
  Let $\cV_{i} = \cV_{i-1} \backslash \cF_{i}$, and $c_i=c_{i-1}$. This
  case is illustrated in Figure~\ref{fig:macFG}(a) for $i=1$.

\item {\bf Case II: $|\cV_{i-1}| < |\cF|$ } 

  Let $\cF_{i} = \cV_{i-1} \cup \cH_{i}$, and $\cV_{i} =\cG \bksl
  \cH_{i}$, and $c_{i}=c_{i-1} + 1$, where $\cH_{i} = \{1,2,\ldots,
  |\cF|-|\cV_{i-1}|\}$. This case is illustrated in
  Figure~\ref{fig:macFG}(b) for $i=4$.
\end{enumerate}
\label{defn:rec}
\end{definition}
To interpret the above construction, we note that the set $\cG$ is a
row-vector with $|\cG|$ elements and let $\cG^{\otimes}$ be obtained
by concatenating $|\cF|$ identical copies of the $\cG$ vector i.e.,
\begin{equation}
\cG^\otimes = \underbrace{\left[ {\cG~|~\cG~|  \ldots  \cG} \right]}_{|\cF| \text{ copies}}
\end{equation}
As shown in Figure~\ref{fig:macFG}, by our construction, the vector
$\cF_1$ spans the first $|\cF|$ elements of $\cG^\otimes$, the vector
$\cF_2$ spans the next $|\cF|$ elements of $\cG^\otimes$ etc. The
constant $c_i$ denotes the index number of copies of the $\cG$ vector
necessary to cover $\cF_i$ .

When $i =
|\cG|$ the row-vector $\cF_i$ terminates exactly at the end of the
last $\cG$ vector in $\cG^\otimes$. Hence, 
\begin{myalign}
  c_{|\cG|} = |\cF| ,\qquad \cV_{|\cG|} = \phi \label{eq:G_params}.
\end{myalign}

By going through the above recursive procedure and invoking
Lemma~\ref{lemma:R1R2_case2} repeatedly, each time by setting $\cF$ in
\eqref{eq:R1bnd_case2_alt} and \eqref{eq:R2bnd_case2_alt} to be
$\cF_i$, we establish the following upper bound on the rate region.
\begin{lemma}
For each $i = 0,1,\ldots, |\cG| $ and the set of channels $\cF_1$,
$\cF_2, \ldots, \cF_{|\cG|}$ defined in Def.~\ref{defn:rec}, the rate
pair $(R_{s,1},R_{s,2})$ satisfies the following upper bound
\begin{myalign}
&i\cdot n (R_{s,1}-\delta_n) + c_i \cdot n (R_{s,2}-\delta_n) \nonumber\\
&\le  {\sum_{j=1}^i} I(M, X_{1,\cB}^n ; Y_{\cF_j}^n) + I(M, X_{1, \cB\backslash \cG }^n; Y_{\cV_i}^n).
\label{eq:RecursiveBound}
\end{myalign}
\label{lem:Recursive}
\end{lemma}
Before providing a proof, we note
that~\eqref{eq:converse_case2_parallel} follows
from~\eqref{eq:RecursiveBound} as described
below. Evaluating~\eqref{eq:RecursiveBound} with $i=|\cG|$, 
using~\eqref{eq:G_params} and letting $\tilde{R}_{s,i}= R_{s,i}-\delta_n$,
\begin{align}
n|\cG| \tilde R_{s,1} + n|\cF| \tilde R_{s,2} &\le \sum_{j=1}^{|\cG|} I(M, X_{1,\cB}^n ; Y_{\cF_j}^n)\\
&= \sum_{j=1}^{|\cG|} \left\{h(Y_{\cF_j}^n)- h(Y_{\cF_j}^n | M, X_{1,\cB}^n)\right\}\\
&= n\left\{|\cG| \cdot |\cF| \cdot \log_2 P + \Theta(1)\right\},\label{eq:R_SumProduct_Bound}
\end{align}
where the last step uses the fact that 
\begin{align}h(Y_{\cF_j}^n) \le \sum_{k\in \cF_j} h(Y_k^n) \le n \{ |\cF| \log_2 P + O(1)\},\end{align} and \begin{align}h(Y_{\cF_j}^n | M, X_{1,\cB}^n) = h(Y_{\cF_j}^n | X_{1, \cF_j}^n, X_{2,\cF_j}^n) =  n\cdot O(1).\end{align} 
Dividing each side of~\eqref{eq:R_SumProduct_Bound} by $\log_2 P$ and
taking the limit $P\rightarrow\infty$
yields~\eqref{eq:converse_case2_parallel}.

\begin{IEEEproof}[Proof of Lemma~\ref{lem:Recursive}]
  We use induction over the variable $i$ to
  establish~\eqref{eq:RecursiveBound}. For $i=0$, note that $c_0=0$
  and $\cV_1 = \cG$ and hence~\eqref{eq:RecursiveBound} is simply
  \eqref{eq:R2bnd_case2_alt}.
This completes the proof for the base case.

For the induction step, we assume that \eqref{eq:RecursiveBound} holds for some $t=i$, we need to show that~\eqref{eq:RecursiveBound} also holds for $t=i+1$, i.e., 
\begin{multline}
  (i+1)\cdot n(R_{s,1}-\delta_n) + c_{i+1} \cdot n(R_{s,2}-\delta_n) \le \\
  \sum_{j=1}^{i+1} I(M, X_{1, \cB}^n; Y_{\cF_j}^n) + I(M, X_{1,
    \cB\bksl \cG}^n; Y_{\cV_{i+1}}^n)
\label{eq:R_induction}
\end{multline} holds. For our proof we separately consider the cases when $|\cF| \le |\cV_i|$ and when $|\cV_i| < |\cF|$ holds.  

When $|\cF| \le |\cV_i|$, from Definition~\ref{defn:rec}
\begin{equation}
\cF_{i+1} \subseteq \cV_i, \qquad \cV_{i+1} = \cV_i \bksl \cF_{i+1}, \qquad c_{i+1} = c_i \label{eq:ind_case_1}
\end{equation}holds.  Then~\eqref{eq:R_induction} follows by combining~\eqref{eq:RecursiveBound} with~\eqref{eq:R1bnd_case2_alt} as we show below. Note that
\begin{align}
& I(M, X_{1, \cB\bksl \cG}^n; Y_{\cV_{i}}^n)=  I(M, X_{1, \cB\bksl \cG}^n; Y_{\cF_{i+1}}^n| Y_{\cV_i \bksl \cF_{i+1}}^n) \notag\\&\qquad + I(M, X_{1, \cB\bksl \cG}^n; Y_{\cV_{i+1}}^n) \label{eq:ind_case_Def_V}\\
&\le  I(M, X_{1, \cB\bksl \cG}^n, Y_{\cV_i \bksl \cF_{i+1}}^n; Y_{\cF_{i+1}}^n ) + I(M, X_{1, \cB\bksl \cG}^n; Y_{\cV_{i+1}}^n)  \\
&\le  I(M, X_{1, \cB\bksl \cG}^n, X_{1, \cV_i \bksl \cF_{i+1}}^n; Y_{\cF_{i+1}}^n ) + I(M, X_{1, \cB\bksl \cG}^n; Y_{\cV_{i+1}}^n)  \label{eq:ind_case_Markov}\\
&\le  I(M, X_{1, \cB\bksl \cG}^n, X_{1, \cG \bksl \cF_{i+1}}^n; Y_{\cF_{i+1}}^n ) + I(M, X_{1, \cB\bksl \cG}^n; Y_{\cV_{i+1}}^n) \label{eq:ind_case_Subset}  \\
&=  I(M, X_{1, \cB\bksl {\cF_{i+1}}}^n ; Y_{\cF_{i+1}}^n ) + I(M, X_{1, \cB\bksl \cG}^n; Y_{\cV_{i+1}}^n)  \label{eq:ind_case_Subset_2}
\end{align}where~\eqref{eq:ind_case_Def_V} follows from the chain rule of the mutual information and the definition of $\cV_{i+1}$ in~\eqref{eq:ind_case_1},
while~\eqref{eq:ind_case_Markov} follows from the Markov condition
\begin{equation}
Y_{\cV_{i}\bksl \cF_{i+1}}^n \leftrightarrow (X_{1,\cV_{i}\bksl \cF_{i+1}}^n, X_{2, \cV_{i} \bksl \cF_{i+1}}^n)\leftrightarrow (M, Y_{\cF_{i+1}}^n, X_{1, \cB\bksl \cG}^n)
\end{equation}and the fact that $M = (X_{2, \cB\cup \cC}^n, Y_{1,\cA}^n)$ already includes $X_{2, \cV_{i} \bksl \cF_{i+1}}^n$,~\eqref{eq:ind_case_Subset} follows from the fact that $\cV_i \subseteq \cG$,
while~\eqref{eq:ind_case_Subset_2} follows from the fact that $\{\cB \bksl \cG\} \cup \{\cG \bksl \cF_{i+1}\} = \{\cB \bksl \cF_{i+1}\}$. 

Substituting~\eqref{eq:ind_case_Subset_2} into the last term in~\eqref{eq:RecursiveBound} we get
\begin{align}
  & i\cdot n ( R_{s,1}-\delta_n) + c_i \cdot n ( R_{s,2}-\delta_n) \nonumber \\
  &\le \sum_{j=1}^i I(M, X_{1, \cB}^n; Y_{\cF_j}^n) + I(M, X_{1, \cB\bksl \cG}^n; Y_{\cV_{i}}^n) \nonumber\\
  &\le \sum_{j=1}^i I(M, X_{1, \cB}^n; Y_{\cF_j}^n) + I(M, X_{1, \cB\bksl
    {\cF_{i+1}}}^n ; Y_{\cF_{i+1}}^n ) \notag \\&\qquad + I(M, X_{1,
    \cB\bksl \cG}^n; Y_{\cV_{i+1}}^n)\label{eq:ind_case1_t1}.
\end{align}Finally combining~\eqref{eq:ind_case1_t1} with~\eqref{eq:R1bnd_case2_alt} and using $c_{i+1}=c_i$ (c.f.~\eqref{eq:ind_case_1}) we have
\begin{align}
  & (i+1)\cdot n (R_{s,1}-\delta_n) + c_{i+1}\cdot n (R_{s,2}-\delta_n) \notag\\
  &\le \sum_{j=1}^i I(M, X_{1, \cB}^n; Y_{\cF_j}^n) + I(M, X_{1, \cB\bksl {\cF_{i+1}}}^n ; Y_{\cF_{i+1}}^n ) \notag \\
  &\qquad + I(M, X_{1, \cB\bksl \cG}^n; Y_{\cV_{i+1}}^n)\nonumber\\
  &\qquad + I(X_{1,\cF_{i+1}}^n; Y_{\cF_{i+1}}^n | M, X_{1, \cB\bksl \cF_{i+1}}^n)\\
  &= \sum_{j=1}^{i+1} I(M, X_{1, \cB}^n; Y_{\cF_j}^n)+ I(M, X_{1,
    \cB\bksl \cG}^n; Y_{\cV_{i+1}}^n)
\end{align} 
as required.

When $|\cF| > |\cV_i|$, as stated in Definition~\ref{defn:rec} we
introduce $\cH_{i+1} = \{1,2,\ldots, |\cF|-|\cV_i|\}$ and recall that
\begin{equation}
\cF_{i+1} = \cV_i \cup \cH_{i+1}, \qquad \cV_{i+1} = \cG\bksl \cH_{i+1}, \qquad c_{i+1} = c_i +1 \label{eq:ind_case_2}
\end{equation}holds.  From~\eqref{eq:R2bnd_case2_alt} and~\eqref{eq:R_induction} we have that
\begin{align}
& i\cdot n (R_{s,1}-\delta_n) + (c_{i}+1) \cdot n (R_{s,2}-\delta_n) \notag \\
&= \sum_{j=1}^i I(M, X_{1, \cB}^n; Y_{\cF_j}^n) + I(M, X_{1, \cB\bksl \cG}^n; Y_{\cV_{i}}^n) \notag\\ &\qquad + I(M, X_{1, \cB\bksl \cG}^n; Y_{\cG}^n)\\
&= \sum_{j=1}^i I(M, X_{1, \cB}^n; Y_{\cF_j}^n) + I(M, X_{1, \cB\bksl \cG}^n; Y_{\cV_{i}}^n) \notag\\&\qquad + I(M, X_{1, \cB\bksl \cG}^n; Y_{\cH_{i+1}}^n|Y_{\cG\bksl \cH_{i+1}}^n)+ I(M, X_{1, \cB\bksl \cG}^n; Y_{\cV_{i+1}}^n)
\label{eq:ind_case2_H_split}
\end{align}

As we will show subsequently,
\begin{multline}
I(M, X_{1, \cB\bksl \cG}^n; Y_{\cV_{i}}^n) + I(M, X_{1, \cB\bksl \cG}^n; Y_{\cH_{i+1}}^n|Y_{\cG\bksl \cH_{i+1}}^n)\\ \le I(M, X_{1, \cB\bksl \cF_{i+1}}^n; Y_{\cF_{i+1}}^n ).
\label{eq:ind_case2_Y_comb}
\end{multline}
Combining~\eqref{eq:R1bnd_case2_alt},~\eqref{eq:ind_case2_H_split} and~\eqref{eq:ind_case2_Y_comb} and using $c_{i+1}=c_i+1$ we get that
\begin{align}
& (i+1)\cdot n (R_{s,1}-\delta_n) + c_{i+1} \cdot n (R_{s,2}-\delta_n) \notag\\
&\le \sum_{j=1}^i I(M, X_{1, \cB}^n; Y_{\cF_j}^n) +  I(M, X_{1, \cB\bksl \cG}^n; Y_{\cV_{i+1}}^n) \notag\\&\quad+ I(M, X_{1, \cB\bksl \cF_{i+1}}^n; Y_{\cF_{i+1}}^n )  + I(X_{\cF_{i+1}}^n ; Y_{\cF_{i+1}}^n|M, X_{1,\cB\bksl \cF_{i+1}}^n)\\
& =\sum_{j=1}^i I(M, X_{1, \cB}^n; Y_{\cF_j}^n) +  I(M, X_{1, \cB\bksl \cG}^n; Y_{\cV_{i+1}}^n)\notag\\&\qquad+ I(M, X_{1, \cB}^n; Y_{\cF_{i+1}}^n ),
\end{align}which establishes~\eqref{eq:R_induction}.

It only remains to establish~\eqref{eq:ind_case2_Y_comb} which we do
now. First, since $\cF_{i+1} \subseteq \cG$ it follows that $\{\cB
\bksl \cG\} \subseteq \{\cB \bksl \cF_{i+1}\}$ and hence we bound the
first term in the left hand side of~\eqref{eq:ind_case2_Y_comb}
as\begin{equation} I(M, X_{1, \cB\bksl \cG}^n; Y_{\cV_{i}}^n) \le I(M,
X_{1, \cB\bksl \cF_{i+1}}^n; Y_{\cV_{i}}^n). \label{eq:ind_comb_part1}
\end{equation} Next, since the set $\cH_{i+1} = \{1,\ldots,
|\cF|-|\cV_i|\}$ constitutes the first $|\cF|-|\cV_i|$ elements of
$\cG$ and $\cV_i = \{|\cG| -|\cV_i|+1,\ldots, |\cG|\}$ constitutes the
last $|\cV_i|$ elements of $\cG$ and $|\cF|\le |\cG|$ we have that
\begin{align}
&\{\cG \bksl \cH_{i+1}\} = \{|\cF|-|\cV_i|+1,\ldots, |\cG|\} \notag\\
&= \{|\cF|-|\cV_i|+1,\ldots, |\cG|-|\cV_i|\} \cup \{|\cG|-|\cV_i|+1 ,\ldots, |\cG|\} \notag\\
&= \{\cG \bksl (\cH_{i+1} \cup \cV_{i})\} \cup \cV_i \notag\\
&= \{\cG \bksl \cF_{i+1}\} \cup \cV_i  \label{eq:HV_split}
\end{align}
where the last relation follows from the definition of $\cF_{i+1}$ (c.f.~\eqref{eq:ind_case_2}).
Using~\eqref{eq:HV_split} we can bound the second term in~\eqref{eq:ind_case2_Y_comb} as follows.
\begin{align}
 &I(M, X_{1, \cB\bksl \cG}^n; Y_{\cH_{i+1}}^n|Y_{\cG\bksl \cH_{i+1}}^n)\notag\\
&= I(M, X_{1, \cB\bksl \cG}^n; Y_{\cH_{i+1}}^n|Y_{\cG\bksl \cF_{i+1}}^n, Y_{\cV_i}^n)\\
&\le I(M, X_{1, \cB\bksl \cG}^n, Y_{G\bksl \cF_{i+1}}^n; Y_{\cH_{i+1}}^n| Y_{\cV_i}^n)\\
&\le I(M, X_{1, \cB\bksl \cG}^n, X_{1, \cG\bksl \cF_{i+1}}^n; Y_{\cH_{i+1}}^n| Y_{\cV_i}^n)\label{eq:ind_Markov}\\
&\le I(M, X_{1, \cB\bksl \cF_{i+1}}^n; Y_{\cH_{i+1}}^n| Y_{\cV_i}^n),\label{eq:ind_comb_part2}
\end{align}
where in~\eqref{eq:ind_Markov}, we use the Markov  relation
\begin{equation}
Y_{\cG\bksl \cF_{i+1}}^n \leftrightarrow (X_{1,\cG\bksl \cF_{i+1}}^n, X_{2,\cG\bksl \cF_{i+1}}^n) \leftrightarrow (M, X_{1,\cB\bksl \cG}^n, Y_{\cF_{i+1}}^n)
\end{equation} 
and the fact that $M = (X_{2, \cB\cup \cC}^n, Y_{1,\cA}^n)$ already contains $X_{2, \cG \bksl \cF_{i+1}}^n$.
Combining~\eqref{eq:ind_comb_part1} and~\eqref{eq:ind_comb_part2} gives
\begin{multline}
 I(M, X_{1, \cB\bksl \cG}^n; Y_{\cV_{i}}^n)  +I(M, X_{1, \cB\bksl \cG}^n; Y_{\cH_{i+1}}^n|Y_{\cG\bksl \cH_{i+1}}^n)\\\le I(M, X_{1, \cB\bksl \cF_{i+1}}^n; Y_{\cF_{i+1}}^n),
\end{multline}
thus establishing~\eqref{eq:ind_case2_Y_comb}.

This completes the proof. 

\subsection{Converse: $\min(|\cA|,|\cC|) \le N_E  \le \max(|\cA|,|\cC|) $}
\label{subsec:converse-parallel-case-3}
We assume without loss of generality that $|\cC| \ge |\cA|$ and as
before let $\cE_k$ be the set of links such that an eavesdropper is
monitoring for message $W_k$. Since $|\cE_1|=|\cE_2|=N_E$ and $|\cA| \le
N_E \le |\cC|$ holds, we select the sets such that the relations $\cA
\subseteq \cE_1 \subseteq \cA \cup \cB$ and $\cC \supseteq \cE_2$ are both
satisfied. Define $\cF=\cB\backslash \cE_1$ and note that
$|\cF|=|\cA|+|\cB|-N_E$.

Theorem~\ref{thm:mainresult} reduces to the following region :
\begin{myalign}
&   0 \le d_1 \le |\cF| \label{eq:converse_case3_R1_parallel}\\
&   0 \le d_2 \le |\cB| + |\cC| - N_E \label{eq:converse_case3_R2_parallel}\\
&   |\cB| d_1+  |\cF| d_2 \le  (|\cB|+|\cC|-N_E) \times |F|  \label{eq:converse_case3_parallel}
\end{myalign}
Since~\eqref{eq:converse_case3_R1_parallel}
and~\eqref{eq:converse_case3_R2_parallel} directly follow from the
single user case~\cite{xiang2010arb}, we only need to
establish~\eqref{eq:converse_case3_parallel}. As in earlier cases we
begin by establishing the following bounds on the rate pair $(R_{s,1},
R_{s,2})$:
  \begin{myalign}
    & n (R_{s,1}-\delta_n) \le I(X_{1,\cF}^n; Y_\cF^n| M, X_{1, \cB\bksl \cF}^n)\label{eq:R1R2_case3_R1_parallel}\\
    & n (R_{s,2}-\delta_n) \le I\left( {M ; Y_\cB^n }\right) + I\left(X_{2, \cC\bksl \cE_2}^n; Y_{\cC \bksl \cE_2}^n
    \right) \label{eq:R1R2_case3_R2_parallel}
  \end{myalign}
  where $M = \left(X_{2, \cB \cup \cC}^n, Y_{1, \cA}^n\right)$. 
\begin{IEEEproof}
The proof for \eqref{eq:R1R2_case3_R1_parallel} is identical to
\eqref{eq:R1bnd_case2_alt} in Lemma~\ref{lemma:R1R2_case2} since the
proof does not depend on the choice of $\cE_2$.  The proof for
\eqref{eq:R1R2_case3_R2_parallel} is identical to
\eqref{eq:case1_R2_bound} in Lemma~\ref{lemma:case1_R1R2_bound}.
\end{IEEEproof}
To
establish~\eqref{eq:converse_case3_R1_parallel}-\eqref{eq:converse_case3_parallel},
note that by defining
\begin{equation}
R_{s,2}' = R_{s,2} - \frac{1}{n}I\left(X_{2, \cC\bksl
      \cE_2}^n; Y_{\cC \bksl \cE_2}^n
  \right),\label{eq:R2_shifted_def}
\end{equation} 
we have from \eqref{eq:R1R2_case3_R2_parallel} that
 \begin{equation}
 n (R_{s,2}'-\delta_n) \le I\left( {M ; Y_\cB^n }\right) \label{eq:R1R2_case3_R2_shifted_parallel}
 \end{equation}
 and the bounds on $R_{s,1}$ and $R_{s,2}'$
 in~\eqref{eq:R1R2_case3_R1_parallel}
 and~\eqref{eq:R1R2_case3_R2_shifted_parallel} are identical to the
 bounds ~\eqref{eq:R1bnd_case2_alt} and~\eqref{eq:R2bnd_case2_alt} in
 Lemma~\ref{lemma:R1R2_case2} with $\cG = \cB$. Applying
 Lemma~\ref{lem:Recursive} to $ R_{s,1}$ and $ R_{s,2}'$ for each
 $i = 0, 1,\cdots, |\cG|$, it follows that
\begin{myalign}
&  i\cdot n (R_{s,1}-\delta_n) + c_i \cdot n (R_{s,2}'-\delta_n) \nonumber\\
& \le  \sum_{j=1}^i I(M, X_{1,\cB}^n;Y_{\cF_j}^n) + I(M,
  Y_{\cV_i}^n).\label{eq:R1R2_case3_recursive}
\end{myalign}
where the sets $\cV_i$, $\cF_i$ and the sequence $c_i$ are
as in
Definition~\ref{defn:rec}. Substituting~\eqref{eq:R1R2_case3_R2_shifted_parallel}
into~\eqref{eq:R1R2_case3_recursive} and evaluating the bound for $i=
|\cB|$ we have that
\begin{multline}
  |\cB| n (R_{s,1}-\delta_n) + |\cF| n ( R_{s,2}-\delta_n) \le |\cF| I( X_{2,\cC\bksl
    \cE_2}^n;Y_{\cC \bksl \cE_2}^n) \\+ \sum_{j=1}^{|\cB|} I(M,
  X_{1,\cB}^n;Y_{\cF_j}^n).
 \label{eq:R1R2_case3_Recursive_End}
\end{multline}
Finally substituting\begin{myalign}
&  d\left( \frac{1}{n}I(M, X_{2, \cC \backslash \cE_2}^n; Y_{\cC\backslash \cE_2}^n\right)  \le |\cC|-N_E\\
&  d\left(\frac{1}{n} I(M, X_{1,\cB}^n;Y_{\cF_j}^n) \right)\le |\cF|,
\end{myalign}
in~\eqref{eq:R1R2_case3_Recursive_End} we obtain~\eqref{eq:converse_case3_parallel}.
 \end{IEEEproof}

\section{General MIMO-MAC}
\label{sec:GeneralMIMO}
The result for the general MIMO case~\eqref{eq:mimomac} follows by a
transformation that reduces the model to the case of parallel
independent channels in the previous section while preserving the
secrecy degrees of freedom region.  As we discuss next, this
transformation involves the generalized singular value decomposition
(GSVD) \cite{paige1981towards} and a channel enhancement argument. For
an analogous application of GSVD to broadcast channels see
e.g.,~\cite{ashish2010secureisit, ersen2010globecom,
  xiang2011globecom}.

\subsection{GSVD Transformation}
\label{sec:GSVD}
\begin{theorem}
\label{thm:GSVD}
\cite{paige1981towards} Given a pair of matrices $\bH_1$ and $\bH_2$
such that the rank of $\bH_i$ is $r_i$, $i=1,2$, and the rank of
$[~\mathbf{H}_1~|~ \mathbf{H}_2~]$ is $r_0$, there exists unitary
matrices $\mathbf{U}_1, \mathbf{U}_2, \mathbf{W}, \mathbf{Q}$ and
nonsingular upper triangular matrix $\mathbf{R}$ such that for
$s=r_1+r_2-r_0$, $\tilde r_1=r_1-s$, $\tilde r_2=r_2-s$,
  \begin{myalign}
& \mathbf{U}_1^H \mathbf{H}^H_1 \mathbf{Q} = \mathbf{\Sigma}_{1 (N_{T_1}
  \times r_0)} \left[ {\mathbf{W}^H \mathbf{R}_{(r_0 \times r_0)},\mathbf{0}}
\right]_{(r_0 \times N_R)} \label{eq:H1_decomp}\\ 
& \mathbf{U}_2^H \mathbf{H}^H_2 \mathbf{Q} = \mathbf{\Sigma}_{2 (N_{T_2}
  \times r_0)} \left[ {\mathbf{W}^H \mathbf{R}_{(r_0 \times r_0)},\mathbf{0}}
\right]_{(r_0 \times N_R)} \label{eq:H2_decomp}\\ 
& \mathbf{\Sigma}_1  = \left[ {\begin{array}{*{20}c}
   {\mathbf{I}_{1 (\tilde r_1 \times \tilde r_1)} } & {} & {}  \\
   {} & {\mathbf{S}_{1 (s \times s)} } & {}  \\
   {} & {} & {\mathbf{O}_{1( (N_{T_1}-\tilde r_1 -s)\times \tilde r_2)} }  \\
\end{array}} \right] \label{eq:Sigma_1}\\
&\mathbf{\Sigma}_2  = \left[ {\begin{array}{*{20}c}
   {\mathbf{O}_{2 ((N_{T_2}- \tilde r_2-s) \times \tilde r_1)}} & {} & {}  \\
   {} & {\mathbf{S}_{2 (s \times s)} } & {}  \\
   {} & {} & {\mathbf{I}_{2 ( \tilde r_2 \times \tilde r_2) } }  \\
\end{array}} \right] \label{eq:Sigma_2}
   \end{myalign}    
   where $\mathbf{I}_i, i=1,2$ are $\tilde r_i \times \tilde r_i$
   identity matrices, $\mathbf{O}_i, i=1,2$ are zero matrices, and $
   {\mathbf{S}_i }, i=1,2$ are $s \times s$ diagonal matrices with
   positive real elements on the diagonal line that satisfy
   $\mathbf{S}_1^2 + \mathbf{S}_2^2 = \bI_s$, and $\tilde r_1 + s+
   \tilde r_2=r_0$. For clarity, the dimension of each matrix is shown
   in the parenthesis in the subscript. $\mathbf{I}_1$ has the same
   number of columns as $\mathbf{O}_2$. $\mathbf{I}_2$ has the same
   number of columns $\mathbf{O}_1$. However, $\mathbf{O}_i, i=1,2$
   are not necessarily square matrices and can be empty, i.e., having
   zero number of rows.
\end{theorem}

For convenience in notation we define $\mathbf{A}=\mathbf{W}^H \mathbf{R}$ and observe
that ${\mathbf{A}}$ is a square and non-singular matrix. Then from
Theorem~\ref{thm:GSVD}, we have:
\begin{myalign}
\mathbf{Q}^H \mathbf{H}_t \mathbf{U}_t = \left[ \begin{array}{l}
 \mathbf{A}^H  \\ 
 \mathbf{0} \\ 
 \end{array} \right]\mathbf{\Sigma}_t^H ,t = 1,2.
\end{myalign}
Without loss of generality, we can cancel $\mathbf{Q}$ and
$\mathbf{U}_t$ and rewrite \eqref{eq:mimomac} as:
\begin{multline}
\mathbf{Y} = \left[ \begin{array}{l}
 \mathbf{A}_{r_0  \times r_0 }^H  \\ 
 \mathbf{0}_{(N_R - r_0) \times r_0} \\ 
 \end{array} \right]_{N_R  \times r_0 } \mathbf{\Sigma}_1^H \mathbf{X}_1  \\ + \left[ \begin{array}{l}
 \mathbf{A}_{r_0  \times r_0 }^H  \\ 
 \mathbf{0}_{(N_R - r_0) \times r_0} \\ 
 \end{array} \right]_{N_R  \times r_0 } \mathbf{\Sigma}_2^H \mathbf{X}_2  + \mathbf{Z}.
\end{multline}
Since $\mathbf{Q}$ and $\mathbf{U}_t$ are unitary matrices, the
components of $\mathbf{Z}$ are independent from each other and the
power constraints of each transmitter remains the same as $\bar P_i,
i=1,2$. Because the components of $\mathbf{Z}$ are independent, the
intended receiver can discard the last $N_R-r_0$ components in
$\mathbf{Y}$ without affecting the secrecy capacity region of this
channel. This means that we only need to consider the case where
$N_R=r_0$ and rewrite \eqref{eq:mimomac} as:
\begin{myalign}
\mathbf{Y} = \mathbf{A}_{r_0  \times r_0 }^H( \mathbf{\Sigma} _1^H
\mathbf{X}_1  +  \mathbf{\Sigma} _2^H \mathbf{X}_2)  + \mathbf{Z}. \label{eq:mimomac2}
\end{myalign}

\subsection{Converse}
\label{subsec:Converse}
For establishing the converse, we further enhance the channel model in
\eqref{eq:mimomac2} to the following
\begin{myalign}
  \mathbf{Y} = \mathbf{\Sigma} _1^H \mathbf{X}_1 + \mathbf{\Sigma}
  _2^H \mathbf{X}_2 + \sigma_+ \mathbf{Z}' \label{eq:mimomac3}
\end{myalign}
where $\sigma_+ \le 1$ is any sufficiently small constant such that,
$\sigma_+^2$ times the maximal eigenvalue of $\mathbf{A}_{r_0 \times
  r_0}^H \mathbf{A}_{r_0 \times r_0}$, is smaller than $1$ and $\mathbf{Z}'$
is a circularly symmetric unit-variance Gaussian noise vector.

To establish~\eqref{eq:mimomac3}, note that we can express  
\begin{equation}\bZ = \sigma_+  \cdot \bA^H \bZ' + \bZ{''}\label{eq:noise_decomp}\end{equation} where 
$\bZ{''}$ is a Gaussian random vector, independent of $\bZ'$ and with a covariance matrix
\begin{myalign}
\mathbf{I}_{r_0 \times r_0} - \sigma_+^2 \mathbf{A}_{r_0 \times r_0}^H \mathbf{A}_{r_0 \times r_0} \label{eq:noiseToDiscard}
\end{myalign}
which is guaranteed to be positive semi-definite by our choice of $\sigma_+$. 
Upon substituting~\eqref{eq:noise_decomp} into~\eqref{eq:mimomac2}, we have
\begin{equation}
  \mathbf{Y} = \mathbf{A}_{r_0 \times r_0 }^H \left( \mathbf{\Sigma} _1^H
  \mathbf{X}_1 + \mathbf{\Sigma} _2^H \mathbf{X}_2 + \sigma_+ \bZ'\right) + \mathbf{Z}{''}.
  \label{eq:mimomac4}
\end{equation}
We consider an enhanced receiver that is revealed $\bZ{''}$. Clearly this additional 
knowledge can only increase the rate and serves as an upper bound. It is also clear that since $\mathbf{Z}{''}$
is independent of $(\bX_1, \bX_2, \bZ')$, it suffices to use this information to cancel $\bZ{''}$ in~\eqref{eq:mimomac4} and 
then discard it. Furthermore since the matrix $\bA$ is invertible, upon canceling it, we obtain~\eqref{eq:mimomac3}.

We further enhance the receiver by replacing ${\mathbf {\Sigma}}_1^H$ and $\mathbf{\Sigma}_2^H$ 
with ${\mathbf {\bar{\Sigma}}}_1^H$ and $\mathbf{\bar{\Sigma}}_2^H$ 
so that the model reduces to
\begin{myalign}
  \mathbf{Y} = \mathbf{\bar{\Sigma}} _1^H \mathbf{X}_1 + \mathbf{\bar{\Sigma}}
  _2^H \mathbf{X}_2 + \sigma_+ \mathbf{Z}' \label{eq:mimomac3e}
\end{myalign}
where \begin{myalign}
&  \mathbf{\bar{\Sigma}}_1^H  = \left[ {\begin{array}{*{20}c}
   {\mathbf{I}_{\tilde r_1  \times \tilde r_1 } } & {} & {}  \\
   {} & {\mathbf{I}_{1(s \times s)} } & {}  \\
   {} & {} & {\mathbf{0}_{1\left( {\tilde r_2  \times \left( {N_{T_1 }  -  r_1} \right)} \right)} }  \\
\end{array}} \right]_{r_0  \times N_{T_1 } } \label{eq:barSigma_1} \\
& \mathbf{\bar{\Sigma}}_2^H  = \left[ {\begin{array}{*{20}c}
   {\mathbf{0}_{2\left( {\tilde r_1  \times \left( {N_{T_2 }  - r_2 } \right)} \right)} } & {} & {}  \\
   {} & {\mathbf{I}_{2(s \times s)} } & {}  \\
   {} & {} & {\mathbf{I}_{2\left( {\tilde r_2  \times \tilde r_2 } \right)} }  \\
\end{array}} \right]_{r_0  \times N_{T_2 } } \label{eq:barSigma_2}
\end{myalign}
are obtained by replacing each diagonal ${\mathbf S}_i$ by the identity matrix. The model~\eqref{eq:mimomac3e} can only have a higher capacity, since each diagonal entry in
${\mathbf{S}}_i$ is between $(0,1)$.  We observe that in the resulting channel model is identical to~\eqref{eq:output_A}-\eqref{eq:output_C} 
\begin{myalign}
&  |\cA|= r_0 - r_2 \label{eq:cA}\\
&  |\cB|= s = r_1+r_2-r_0 \label{eq:cB}\\
&  |\cC|= r_0 - r_1 \label{eq:cC}
\end{myalign} except that the noise variance is reduced by a factor of
$\sigma_+^2$.  Since a fixed scaling in the noise power does not
affect the secure-degrees of freedom, an outer bound on the
s.d.o.f. for the parallel channel
model~\eqref{eq:output_A}-\eqref{eq:output_C} with $\cA$, $\cB$ and
$\cC$ defined via~\eqref{eq:mimomac3e}, continues to be an outer bound
on the s.d.o.f. region for the general MIMO-MAC channel.

Substituting~\eqref{eq:cA}-\eqref{eq:cC} in the upper bounds in section \ref{subsec:converse-parallel-case-1},
\ref{subsec:converse-parallel-case-2} and \ref{subsec:converse-parallel-case-3}
we establish the converse in Theorem~\ref{thm:mainresult}.

\subsection{Achievability}
\label{subsec:achiev}
To establish the achievability for the general MIMO case we further
use a suitable degradation mechanism to reduce the
model~\eqref{eq:mimomac2} to
\begin{myalign}
  \mathbf{Y} = \mathbf{\Sigma} _1^H \mathbf{X}_1 + \mathbf{\Sigma}
  _2^H \mathbf{X}_2 + \sigma \mathbf{Z}'' \label{eq:mimomac5}
\end{myalign}
where $\sigma \ge 1$ is any sufficiently large constant such that,
$\sigma^2$ times the minimum eigenvalue of $\mathbf{A}_{r_0 \times
  r_0}^H \mathbf{A}_{r_0 \times r_0}$, is greater than $1$ and $\mathbf{Z}''$
is a circularly symmetric unit-variance Gaussian noise vector.
Since $\bA$ is non-singular we are guaranteed that all the singular values of $\bA$
are non-zero and hence a $\sigma < \infty$ exists. 

To establish~\eqref{eq:mimomac5},  let $\bZ'$ be a Gaussian
noise vector with covariance
\begin{myalign}
\sigma ^2 \mathbf{A}_{r_0 \times r_0}^H \mathbf{A}_{r_0 \times r_0}- \mathbf{I}_{r_0 \times r_0}  \label{eq:noiseToAdd}
\end{myalign} independent of $\bZ$
and consider a degraded version of~\eqref{eq:mimomac2}
\begin{equation}
  \mathbf{Y} = \mathbf{A}_{r_0 \times r_0 }^H \left( \mathbf{\Sigma} _1^H
  \mathbf{X}_1 + \mathbf{\Sigma} _2^H \mathbf{X}_2\right) + \bZ + \bZ'
  \label{eq:mimomac4b}
\end{equation}
which can be simulated at the receiver by adding additional noise
$\bZ'$ to its output. Since $\bZ + \bZ' \sim {\mathcal{CN}}(0,
\sigma^2 \bA^H \bA )$, we can express $\bZ + \bZ' =\sigma \bA^H
\bZ''$. Substituting into~\eqref{eq:mimomac4b} and canceling the
non-singular matrix $\bA$, we arrive at~\eqref{eq:mimomac5}.

Let $\bar{s}>0$ denote the minimum element on the diagonals of $\bS_1$ and $\bS_2$ in~\eqref{eq:Sigma_1}
and~\eqref{eq:Sigma_2} respectively. By appropriately scaling down the transmit powers on each of the sub-channels 
we can further reduce~\eqref{eq:mimomac4} to
\begin{equation}
  \mathbf{Y} =  \mathbf{\bar{\Sigma}} _1^H
  \mathbf{X}_1 + \mathbf{\bar{\Sigma}} _2^H \mathbf{X}_2 + \frac{\sigma}{\bar{s}} \bZ''
  \label{eq:mimomac6}
\end{equation}
where $\mathbf{\bar{\Sigma}}_k$ are defined in~\eqref{eq:Sigma_1} and~\eqref{eq:Sigma_2} respectively. 
The model~\eqref{eq:mimomac6} is identical to the parallel channel model~\eqref{eq:output_A}-\eqref{eq:output_C} with the size of 
sets $\cA$, $\cB$ and $\cC$ in~\eqref{eq:cA}-\eqref{eq:cC} and with a noise power that is larger by a factor of $\sigma^2/{\bar{s}^2}$.
Since a constant factor in the noise power does not affect the secrecy degrees of freedom, the coding schemes described in
section~\ref{subsec:achievable-parallel} achieves the lower bound in Theorem~\ref{thm:mainresult}.

\section{Conclusion}
\label{sec:conclusion}
In this work we have studied the two-transmitter Gaussian complex
MIMO-MAC wiretap channel where the eavesdropper channel is arbitrarily
varying and its state is known to the eavesdropper only, and the main
channel is static and its state is known to all nodes. We have
completely characterized the s.d.o.f. region for this channel for all
possible antenna configurations. We have proved that this
s.d.o.f. region can be achieved by a scheme that orthogonalizes the
transmit signals of the two users at the intended receiver, in which
each user achieves secrecy guarantee independently without cooperation
from the other user. The converse was proved by carefully changing the
set of signals available to the eavesdropper through an induction
procedure in order to obtain an upper bound on a weighted-sum-rate
expression.

As suggested by this work, the optimal strategy for a communication
network where the eavesdropper channel is arbitrarily varying can
potentially be very different from the case where the eavesdropper
channel is fixed and its state is known to all terminals. This is also
observed for example in the MIMO broadcast channel
\cite{xiang2011globecom} and the two-way channel \cite{xiang2011isit,
  xiang2011globecomworkshop}. Characterizing secure transmission
limits for a broader class of communication models with this
assumption is hence important and is left as future work.

\appendices

\section{Proof of Lemma~\ref{lemma:case1_R1R2_bound}}
\label{app:R1R2_case1}

For $R_{s,1}$, from Fano's inequality, we have
\begin{myalign}
 n(R_{s,1}-\delta_n) & \le I(W_1; Y_{\cA \cup \cB \cup \cC}^n)- I(W_1; Y_{\cE_1}^n)\\
 & \le I\left( {W_1 ;Y_{\cA \cup \cB \cup C}^n |Y_{\cE_1 }^n
   } \right) \\ 
&  \le I\left( {W_1 ;Y_{\cA \cup \cB \cup \cC}^n ,X_{2,\cB \cup \cC}^n |Y_{\cE_1 }^n }
\right) \\
&  = I\left( {W_1 ;Y_{\cA \cup \cB}^n ,X_{2,\cB \cup \cC}^n |Y_{\cE_1 }^n } \right)  \label{eq:beforemoveX2C}
\end{myalign}
where the last step~\eqref{eq:beforemoveX2C} relies on the fact that the additive noise at each
receiver end of each sub-channel in Figure~\ref{fig:setA,B,C} is
independent from each other and hence
$$Y_{\cC}^n \rightarrow X_{2,\cC}^n \rightarrow (W_1, Y_{\cA \cup \cB}^n, Y_{\cE_1}^n, X_{2, \cB}^n)$$ holds. 
Since $\left(X_{2,\cC}^n, X_{2, \cB}^n\right)$ is independent from
$W_1$, and $\cE_1 \subseteq \cA$, \eqref{eq:beforemoveX2C} can be written as:
\begin{myalign}
& I\left( {W_1 ;Y_{\cA \cup \cB}^n |Y_{\cE_1 }^n, X_{2, \cB\cup\cC}^n  } \right) \notag\\ 
  = &I\left( {W_1 ;Y_{\left( {\cA\backslash \cE_1 } \right) \cup \cB}^n |Y_{\cE_1 }^n ,X_{2,\cB \cup \cC}^n } \right) \\ 
  = &I\left( {W_1 ;Y_{\cA\backslash \cE_1 }^n |Y_{\cE_1 }^n ,X_{2,\cB \cup \cC}^n } \right) + I\left( {W_1 ;Y_\cB^n |Y_\cA^n ,X_{2,\cB \cup \cC}^n } \right) \label{eq:R1_case1_split}
  \end{myalign}where the last step~\eqref{eq:R1_case1_split} follows from the fact $\cE_1 \subseteq \cA$ and hence $\cA = (\cA \bksl \cE_1) \cup \cE_1$.
  We separately bound each of the two terms above.
\begin{myalign}
  &I\left( {W_1 ;Y_{\cA\backslash \cE_1 }^n |Y_{\cE_1 }^n ,X_{2,\cB \cup \cC}^n } \right) \notag\\
  &\le I\left( {W_1, Y_{\cE_1 }^n ,X_{2,\cB \cup \cC}^n } ;Y_{\cA\backslash \cE_1 }^n  \right) \\
  &\le I\left( {W_1, Y_{\cE_1 }^n ,X_{2,\cB \cup \cC}^n }, X_{1, \cA \bksl \cE_1}^n ;Y_{\cA\backslash \cE_1 }^n  \right) \\  
    &= I\left( X_{1, \cA \bksl \cE_1}^n ;Y_{\cA\backslash \cE_1 }^n  \right)  \label{eq:R1_case1_split_part1}
  \end{myalign}  
  where the last step follows from the Markov chain relation $Y_{1, \cA \bksl \cE_1}^n \leftrightarrow  X_{\cA\backslash \cE_1 }^n  \leftrightarrow (W_1, Y_{\cE_1 }^n ,X_{2,\cB \cup \cC}^n )$, 
  We upper bound the second term in~\eqref{eq:R1_case1_split} as follows

  \begin{myalign}
 & I\left( {W_1 ;Y_\cB^n |Y_\cA^n ,X_{2,\cB \cup \cC}^n } \right)\notag\\
&\le I\left( {X^n_{1,\cA\cup\cB} ;Y_\cB^n |Y_\cA^n ,X_{2,\cB \cup \cC}^n } \right) \label{eq:R1_case1_W1_markov}\\
&= I\left( {X^n_{1,\cB} ;Y_\cB^n |Y_\cA^n ,X_{2,\cB \cup \cC}^n } \right)  \notag\\ &\quad + I\left( {X^n_{1,\cA} ;Y_\cB^n |Y_\cA^n ,X_{2,\cB \cup \cC}^n }, X_{1,\cB}^n \right)   \label{eq:R1_case1_Y_markov}\\
&= I\left( {X^n_{1,\cB} ;Y_\cB^n |Y_\cA^n ,X_{2,\cB \cup \cC}^n } \right) \label{eq:R1_case1_split_part2}\end{myalign}  
where we use the Markov relation $W_1 \leftrightarrow X^n_{1, \cA \cup \cB} \leftrightarrow (Y_\cA^n ,X_{2,\cB \cup \cC}^n)$ in 
step \eqref{eq:R1_case1_W1_markov} and~\eqref{eq:R1_case1_split_part2} follows from the fact Markov relation \begin{equation}
Y_{\cB}^n \leftrightarrow (X_{1,\cB}^n, X_{2,\cB}^n) \leftrightarrow (X_{2,\cC}^n, Y_{\cA}^n).
\end{equation}
Note that~\eqref{eq:case1_R1_bound} follows upon substituting~\eqref{eq:R1_case1_split_part1} and~\eqref{eq:R1_case1_split_part2} into~\eqref{eq:R1_case1_split}.

For $R_{s,2}$, from Fano's inequality and the secrecy constraint, we have:
\begin{myalign}
& n(R_{s,2}-\delta_n) \le  I(W_2; Y_{\cA \cup \cB \cup \cC}^n)-I(W_2; X_{2,\cE_2 }^n)\\
\le& I\left( {W_2 ;Y_{\cA \cup \cB \cup \cC}^n |X_{2,\cE_2 }^n } \right) \\ 
  =& I\left( {W_2 ;Y_{\cB \cup \cC}^n |Y_\cA^n ,X_{2,\cE_2 }^n } \right) \label{eq:case1_R2_YA_Markov} \\ 
  =& I\left( {W_2 ;Y_{(\cC\backslash \cE_2 ) \cup \cB}^n |Y_\cA^n ,X_{2,\cE_2 }^n } \right) \label{eq:case1_R2_YE_Markov} \\ 
  =& I\left( {W_2 ;Y_{\cC\backslash \cE_2 }^n |Y_\cA^n ,X_{2,\cE_2 }^n } \right)
+ I\left( {W_2 ;Y_\cB^n |Y_\cA^n ,X_{2,\cE_2 }^n ,Y_{\cC\backslash \cE_2 }^n }
\right) \label{eq:R2_case1_split} 
\end{myalign}
where~\eqref{eq:case1_R2_YA_Markov} follows from the fact that $Y_\cA^n$ is independent of $(W_2, X_{2, \cB\cup\cC}^n)$
and~\eqref{eq:case1_R2_YE_Markov} follows from the fact that $Y_{\cE_2}^n \rightarrow X_{2,\cE_2}^n \rightarrow (Y_{\cB \cup \cC\bksl \cE_2}^n, W_2, Y_{\cA}^n)$
holds. We separately bound each term in~\eqref{eq:R2_case1_split}.
\begin{myalign}
& I\left( {W_2 ;Y_{\cC\backslash \cE_2 }^n |Y_\cA^n ,X_{2,\cE_2 }^n } \right) \le I\left( {W_2, Y_\cA^n ,X_{2,\cE_2 }^n ;Y_{\cC\backslash \cE_2 }^n  } \right)\\
&\le I\left( X_{2,\cC\bksl \cE_2}^n, {W_2, Y_\cA^n ,X_{2,\cE_2 }^n ;Y_{\cC\backslash \cE_2 }^n  } \right)\\
&= I\left( X_{2,\cC\bksl \cE_2}^n ;Y_{\cC\backslash \cE_2 }^n   \right), \label{eq:R2_case1_split_part1}
\end{myalign} where the justification for establishing~\eqref{eq:R2_case1_split_part1} is identical to~\eqref{eq:R1_case1_split_part1}
and hence omitted. We  finally bound the second term in~\eqref{eq:R2_case1_split}.
\begin{myalign}
 & I\left( {W_2 ;Y_\cB^n |Y_\cA^n ,X_{2,\cE_2 }^n ,Y_{C\backslash \cE_2 }^n } \right) \\ 
  \le&  I\left( {X_{2,\cB \cup \cC}^n ;Y_\cB^n |Y_\cA^n ,X_{2,\cE_2 }^n ,Y_{\cC\backslash \cE_2 }^n } \right) \\ 
  \le&  I\left( {Y_\cA^n ,X_{2,\cB \cup \cC}^n ,X_{2,\cE_2 }^n,
      Y_{\cC\backslash \cE_2 }^n ;Y_\cB^n } \right) \\ 
  =& I\left( {Y_\cA^n ,X_{2,\cB \cup \cC}^n ,X_{2,\cE_2 }^n ;Y_\cB^n }
  \right)\nonumber \\ 
  &+ I\left( { Y_{\cC\backslash \cE_2 }^n ;Y_\cB^n |Y_\cA^n ,X_{2,\cB \cup \cC}^n ,X_{2,\cE_2 }^n} \right)\\
  =&  I\left( {Y_\cA^n ,X_{2,\cB \cup \cC}^n ;Y_\cB^n } \right)   \label{eq:R2_case1_split_part2}
\end{myalign}where the justification for arriving at~\eqref{eq:R2_case1_split_part2} is similar to~\eqref{eq:R1_case1_split_part2}
and hence omitted.

Substituting~\eqref{eq:R2_case1_split_part1} and~\eqref{eq:R2_case1_split_part2} into~\eqref{eq:R2_case1_split}
we establish~\eqref{eq:case1_R2_bound}. 

\section{Proof of Lemma~\ref{lemma:R1R2_case2}}
\label{app:R1R2_case2}

Assume the eavesdropper monitors $Y_\cA^n$ and $X_{1,\cE_1\backslash \cA}^n$
for $W_1$. Then for $R_{s,1}$, from Fano's inequality, we have:
\begin{myalign}
& n(R_{s,1}-\delta_n)\notag\\
  \le& I\left( {W_1 ;Y_{\cA \cup \cB \cup \cC}^n }\right)- I\left({W_1; Y_\cA^n ,X_{1,\cE_1 \backslash \cA}^n } \right) \\ 
  \le& I\left( {W_1 ;Y_{\cA \cup \cB \cup \cC}^n |Y_\cA^n ,X_{1,\cE_1 \backslash \cA}^n } \right) \\ 
  =& I\left( {W_1 ;Y_{ \cB \cup \cC}^n |Y_\cA^n ,X_{1,\cE_1 \backslash \cA}^n } \right) \\ 
  \le& I\left( {W_1 ;Y_{\cB \cup \cC}^n ,X_{2,\cB \cup \cC}^n |Y_\cA^n ,X_{1,\cE_1 \backslash \cA}^n } \right) \\ 
  =&I\left( {W_1 ;Y_{\cB \cup \cC}^n  |Y_\cA^n ,X_{1,\cE_1 \backslash \cA}^n }, X_{2,\cB \cup \cC}^n \right) \label{eq:R1_case2_X2BC_indep}\\ 
  =&I\left( {W_1 ;Y_{\cF }^n  |Y_\cA^n ,X_{1,\cE_1 \backslash \cA}^n }, X_{2,\cB \cup \cC}^n \right) \label{eq:R1_case2_F_Markov}\\ 
  =&I\left( {W_1 ;Y_{\cF }^n  |Y_\cA^n ,X_{1,\cB \backslash \cF}^n }, X_{2,\cB \cup \cC}^n \right) \label{eq:R1_case2_step1} 
  \end{myalign}
     where~\eqref{eq:R1_case2_X2BC_indep} follows from the fact that $X_{2, \cB\cup \cC}^n$ is independent of $(W_1,Y_\cA^n,X_{1, \cE_1\bksl \cA}^n)$.
 while~\eqref{eq:R1_case2_F_Markov} follows from the fact that since the noise across the channels is independent the Markov condition 
 $$(Y_{ \cE_1 \bksl\cA}^n, Y_{\cC}^n) \leftrightarrow (X_{1, \cE_1 \bksl\cA}^n, X_{2,\cB\cup \cC}^n)  \leftrightarrow (W_1, Y_{B\bksl \cE_1}^n,Y_{\cA}^n)$$
holds and furthermore we have defined $\cF = \cB \bksl \cE_1$.

Since the channel noise is independent of the message, $W_1 \leftrightarrow X_{1, \cA \cup \cB}^n \leftrightarrow (Y_{\cF \cup \cA}^n, X_{1, \cB\bksl\cF}^n, X_{2, \cB \cup \cC}^n)$
holds. Hence
\begin{myalign}     
& I\left( {W_1 ;Y_{\cF }^n  |Y_\cA^n ,X_{1,\cB \backslash \cF}^n }, X_{2,\cB \cup \cC}^n \right)\\
\le & I\left( {X_{1, \cA \cup \cB}^n ;Y_{\cF }^n  |Y_\cA^n ,X_{1,\cB \backslash \cF}^n }, X_{2,\cB \cup \cC}^n \right)\\
= & I\left( {X_{1, \cF}^n ;Y_{\cF }^n  |Y_\cA^n ,X_{1,\cB \backslash \cF}^n }, X_{2,\cB \cup \cC}^n \right)\notag\\
&\qquad + I\left( {X_{1, \cA \cup \cB \bksl \cF}^n ;Y_{\cF }^n  |Y_\cA^n ,X_{1,\cB }^n }, X_{2,\cB \cup \cC}^n \right)\label{eq:R1_case2_YF_Markov}\\
= & I\left( {X_{1, \cF}^n ;Y_{\cF }^n  |Y_\cA^n ,X_{1,\cB \backslash \cF}^n }, X_{2,\cB \cup \cC}^n \right)\label{eq:R1_case2_step2} 
\end{myalign}
where the last step uses the fact that the second term in~\eqref{eq:R1_case2_YF_Markov} involves conditioning on $(X_{1,\cF}^n, X_{2,\cF}^n)$ and hence is zero.
 This establishes~\eqref{eq:R1bnd_case2_alt}.

For $R_{s,2}$, we assume the eavesdropper is
monitoring $ X_{2,\cC}^n ,X_{2,\cE_2 \backslash \cC}^n$ for $W_2$. Using
Fano's inequality and the secrecy constraint, we have:

\begin{myalign}
 &n(R_{s,2}-\delta_n) 
 \le I\left( W_2 ;Y_{\cA \cup \cB \cup \cC}^n\right)-I\left(W_2; X_{2,\cE_2}^n  \right) \\ 
 \le& I\left( {W_2 ;Y_{\cA \cup \cB \cup \cC}^n} |X_{2,\cE_2}^n  \right) \\ 
  \le& I\left( {W_2 ;Y_{\cA \cup \cB \cup \cC}^n}, X_{1, \cE_2 \cap \cB}^n |X_{2,\cE_2}^n  \right) \\ 
   =& I\left( W_2 ;Y_{\cB \cup \cC}^n |X_{2,\cE_2}^n, Y_{\cA}^n  , X_{1, \cE_2 \cap \cB}^n \right) \label{eq:R2_case2_Tx1_indep}\\ 
  \le& I\left( {X_{2,\cB \cup \cC}^n ;Y_{ \cB \cup \cC}^n |X_{2,\cE_2}^n, Y_{\cA}^n  },X_{1, \cE_2 \cap \cB}^n \right) \\ 
  =& I\left( {X_{2,\cB \cup \cC}^n ;Y_{ \cG \cup \cE_2}^n |X_{2,\cE_2}^n, Y_{\cA}^n  },X_{1, \cE_2 \cap \cB}^n \right) \label{eq:R2_case2_EG_union}\\ 
   =& I\left( {X_{2,\cB \cup \cC}^n ;Y_{ \cG }^n |X_{2,\cE_2}^n,
       Y_{\cA}^n, X_{1, \cE_2 \cap \cB}^n  } \right) \notag\\&\qquad+
   I\left( {X_{2,\cB \cup \cC}^n ;Y_{ \cE_2 }^n |X_{2,\cE_2}^n, Y_{\cA
         \cup \cG}^n, X_{1, \cE_2 \cap \cB}^n  } \right) \label{eq:R2_case2_Tx1_YE_Markov_split}\\ 
   =& I\left( {X_{2,\cB \cup \cC}^n ;Y_{ \cG }^n |X_{2,\cE_2}^n, Y_{\cA}^n,X_{1, \cE_2 \cap \cB}^n  } \right)\label{eq:R2_case2_Tx1_YE_Markov}\\
  \le&\ I\left( X_{2,\cB \cup \cC}^n, Y_{\cA}^n,X_{1, \cE_2 \cap \cB}^n ;Y_{ \cG}^n \right) \\ 
  \le& I\left( M, X_{1,\cB\bksl \cG}^n ;Y_{ \cG}^n \right) 
\end{myalign}where~\eqref{eq:R2_case2_Tx1_indep} follows from the fact
that $(X_{1, \cE_2\cap \cB}^n, Y_{\cA}^n)$ are the transmitted signals
from user 1 and independent of $(W_2, X_{2,\cE_2}^n)$
and~\eqref{eq:R2_case2_EG_union} follows from the fact that $\cC
\subseteq \cE_2 \subseteq \cB\cup \cC$ and $\cG = \cB \bksl\cE_2$ and
hence $\cE_2 \cup \cG = \cB \cup \cC$
holds. Eq.~\eqref{eq:R2_case2_Tx1_YE_Markov} follows from the fact
that since the noise on each channel is Markov, we have $Y_{\cE_2}^n
\leftrightarrow (X_{2,\cE_2}^n, X_{1, \cE_2 \cap \cB}^n)
\leftrightarrow (Y_{\cA \cup \cG}^n, X_{\cB\cup\cC}^n) $ and hence the
second term in \eqref{eq:R2_case2_Tx1_YE_Markov_split} is zero.

Hence we have proved Lemma~\ref{lemma:R1R2_case2}.

\bibliographystyle{IEEEtran}

\end{document}